\documentclass{article}
\usepackage[utf8]{inputenc}
\usepackage[utf8]{inputenc}
\usepackage[T1]{fontenc}
\usepackage[francais]{babel}
\usepackage{graphicx}
\usepackage{eurosym}
\usepackage{geometry}
\usepackage{mathtools}
\usepackage{float}
\usepackage{multirow}
\usepackage{amsmath,amssymb}
\usepackage{a4wide}
\usepackage[bookmarksopen,bookmarksdepth=2,breaklinks=true]{hyperref}
\usepackage{placeins}
\usepackage{geometry}
\usepackage{eso-pic}
\usepackage{bbold}
\usepackage{natbib}
\newcommand{\beginsupplement}{%
        \setcounter{table}{0}
        \renewcommand{\thetable}{S\arabic{table}}%
        \setcounter{figure}{0}
        \renewcommand{\thefigure}{S\arabic{figure}}%
     }

\title{A location-scale joint model for studying the link between the time-dependent subject-specific variability of blood pressure and competing events}

\author
{Léonie Courcoul$^{1*}$, Christophe Tzourio$^1$, Mark Woodward$^{2,3}$,\\
Antoine Barbieri$^1$,  and Hélène Jacqmin-Gadda$^1$}
\date{}
\begin{document}

\maketitle
\begin{center}
\noindent$^1$Univ. Bordeaux, INSERM, Bordeaux Population Health, U1219, France\\
$^2$ The George Institute for Global Health, Imperial College London, UK\\
$^3$ The George Institute for Global Health, University of New South Wales, Sydney, Australia\\
\end{center}

\begin{abstract}
Given the high incidence of cardio and cerebrovascular diseases (CVD), and its association with morbidity and mortality, its prevention is a major public health issue. A high level of blood pressure is a well-known risk factor for these events and an increasing number of studies suggest that blood pressure variability may also be an independent risk factor. However, these studies suffer from significant methodological weaknesses. In this work we propose a new location-scale joint model for the repeated measures of a marker and competing events. This joint model combines a mixed model including a subject-specific and time-dependent residual variance modeled through random effects, and cause-specific proportional intensity models for the competing events. The risk of events may depend simultaneously on the current value of the variance, as well as, the current value and the current slope of the marker trajectory. The model is estimated by maximizing the likelihood function using the Marquardt-Levenberg algorithm. The estimation procedure is implemented in an R-package and is validated through a simulation study. This model is applied to study the association between blood pressure variability and the risk of CVD and death from other causes. Using data from a large clinical trial on the secondary prevention of stroke, we find that the current individual variability of blood pressure is associated with the risk of CVD and death. Moreover, the comparison with a model without heterogeneous variance shows the importance of taking into account this variability in the goodness-of-fit and for dynamic predictions.
\end{abstract}


\maketitle

\section{Introduction}
Cardiovascular diseases, such as ischaemic heart disease, and cerebrovascular events are two leading causes of death. Moreover these diseases lead often to acquired physical disability or to dementia. In addition, medical care and disability management following this type of disease generate significant societal, human, and financial distress \citep{de_pouvourville_2016}. Given the frequency of cardio and cerebrovascular diseases (CVD) and its dramatic consequences at the individual and societal level, the identification of modifiable risk factors is essential to implement prevention programs. Hypertension (high values of blood pressure) is a long-known major risk factor for these diseases. The prevalence of hypertension is high, increases with age, and effective blood pressure-lowering treatments are available. More recently, the visit-to-visit variability of blood pressure has been shown to be associated with an increased risk of stroke and cardiovascular events independently of the level of blood pressure in several studies \citep{pringle_2003,rothwell_2010,shimbo_2012}.

Most of the previous studies have used the individual empirical standard deviation, or some other measure of variation (e.g. the coefficient of variation) or extreme value (e.g. the maximum), of blood pressure as an explanatory variable in a Cox model for the event risk. However, they were exposed to methodological issues. A first strategy consists of calculating the empirical standard deviation of blood pressure on all available measurements \citep{mehlum_2018}. This strategy induces conditioning on the future, likely leading to bias because measurements after the current time (and sometimes after the event time) are used to predict the event at the current time \citep{andersen_2012, decourson_2021}. A second strategy consists of computing the standard deviation of blood pressure on the measurements collected over an initial period of the study, keeping in the sample only the individuals who did not have the event before the end of this period in order to predict the risk beyond this period \citep{decourson_2021}. This could induce selection bias and certainly creates loss of power. To avoid these issues, the standard deviation of blood pressure can be considered as a time-dependent variable and calculated using only measurements before the event. Nevertheless, this approach neglects the measurement error of the standard deviation, which is a serious issue when the number of measurements differs between individuals, and requires imputation of the standard deviation at all event times. These limitations may introduce bias \citep{prentice_1982}. Moreover, blood pressure and its standard deviation are endogenous variables, and the Cox model is not adapted to this type of variable \citep{commenges_2015}. Finally, it is essential to account for competing death from other causes because mortality and CVD risk both increases with age and may be both associated with blood pressure. 

Joint models allow simultaneous analysis of longitudinal data and clinical events. They combine a mixed model for repeated measures of exposure and a time-to-event model. Functions of the random effects from the mixed model are included as explanatory variables in the time-to-event model to account for the association between the two outcomes. This allows evaluation of the impact of the longitudinal data on the event risk without bias, contrary to the two stage estimation \citep{rizopoulos_2012, tsiatis_2004,henderson_2000}.

Location-scale mixed models have been introduced to investigate the heterogeneity of intra-subject variability for longitudinal data by introducing random effects in the variance modelling \citep{hedecker_2013}. For studying the association between the variability of a biomarker and a clinical event, Gao et al.\citep{gao_2011} and Barrett et al.\citep{barrett_2019} have proposed a joint model combining a mixed model including a subject-specific random effect for the residual variance and a proportional hazard model for the event risk. However, the considered dependence structure is quite restrictive since, in their models, the event risk depends only on the random effects and not on time-dependent characteristics of the marker trajectory, such as the current value or the current slope. In addition, none of them assumes for time-dependent subject-specific variability of the maker and they do not handle competing events.

The objective of our work was, therefore, to propose a new location-scale joint model accounting for both time-dependent individual variability of a marker and competing events. To do this, we extended the model proposed by Gao et al.\citep{gao_2011} and Barrett et al.\citep{barrett_2019} to include a time-dependent variability, competing events, a more flexible dependence structure between the event and the marker trajectory, and more flexible baseline risk functions. In contrast to the previous works we propose a frequentist estimation approach which is implemented in the R-package \texttt{FlexVarJM}.

This paper is organized as follows. Section 2 describes the model and the estimation procedure using a robust algorithm for maximizing the likelihood. Section 3 presents a simulation study to assess the estimation procedure performance. In section 4, the model is applied to the data from the Perindopril Protection Against Stroke Study (PROGRESS) clinical trial, a blood-pressure lowering trial for the secondary prevention of stroke  \citep{progress_2001}. Finally, Section 5 concludes this work with some elements of discussion.

\section{Method}\label{sec2}
Let us consider a sample of $N$ individuals. For each individual $i \in \{1,...,N\}$, we consider the $n_i$-vector of repeated measures $Y_i = (Y_{i1},...,Y_{in_i})^\top$ with $Y_{ij}$ the value of the longitudinal outcome of individual $i$ at time $t_{ij}$ $(j = 1,\ldots,n_i)$. Assuming two competing events, we denote $T_i = \min(T^*_{i1},T^*_{i2},C_i)$ the observed time with $T^*_{ik}$ the real time for the event $k$ $(k=1,2)$ and $C_i$ the censoring time for the $i$th individual. Censoring event and real time are supposed to be independent. We then denote $\delta_i \in \{0,1,2\}$ the individual event indicator such as $\delta_i = k$ if the competing event $k \in \{1,2\}$ occurs and $\delta_i = 0$ otherwise.

\subsection{Joint model with time-dependent individual variability}\label{subsec1}
We propose joint modelling for a longitudinal outcome and competing events using a shared random-effect approach. The longitudinal submodel is defined by a linear mixed-effect model with heterogeneous variance:
\begin{equation}
\left\{
    \begin{array}{ll}
         Y_{ij} = Y_{i}(t_{ij}) = \widetilde{Y}_i(t_{ij}) + \epsilon_{ij} = X_{ij}^{\top} \beta+Z_{ij}^{\top} b_{i}+\epsilon_{i}(t_{ij}), \\
        \epsilon_{ij}(t_{ij}) \sim \mathcal{N}(0,\sigma_i^2(t_{ij})) \hspace{3mm} \text{with} \hspace{3mm} \log(\sigma_i(t_{ij}))  = O_{ij}^{\top} \mu+M_{ij}^{\top} \tau_{i}
    \end{array}
\right.  
 \label{Mixed}
\end{equation}
with $X_{ij}$, $O_{ij}$, $Z_{ij}$ and $M_{ij}$ four vectors of explanatory variables for subject $i$ at visit $j$, respectively associated with the fixed-effect vectors $\beta$ and $\mu$, and the subject-specific random-effect vector $b_i$ and $\tau_i$, such as 
$$\quad\left(\begin{array}{c}
b_{i} \\
\tau_i
\end{array}\right) \sim \mathcal{N}\left(\left(\begin{array}{c}
0 \\
0
\end{array}\right),\left(\begin{array}{cc}
\Sigma_{b} & \Sigma_{\tau b} \\
\Sigma_{\tau b}^{\top} & \Sigma_{\tau}
\end{array}\right)\right)$$

\noindent The risk function for the event $k \in \{1,2\}$ is defined by:
\begin{equation}
    \lambda_{ik}(t)=\lambda_{0k}(t) \exp \left(W_{i}^{\top} \gamma_{k}+\alpha_{1k}\tilde{y}_i(t)+\\
     \alpha_{2k}\tilde{y}'_i(t)+ \alpha_{\sigma k} \sigma_i(t) \right),
    \label{Surv}
\end{equation}
with $\lambda_{0k}(t)$ the baseline risk function, $W_{i}$ a vector of baseline covariates associated with the regression coefficient $\gamma_k$, and $\alpha_{1k}$, $\alpha_{2k}$ and $\alpha_{\sigma k}$ the regression coefficients associated with the current value $\tilde{y}_i(t)$, the current slope $\tilde{y}'_i(t)$ and the current variability $\sigma_i(t)$ of the marker, respectively. Different parametric forms for the baseline risk function can be considered, such as exponential, Weibull, or, for more flexibility, a B-splines base with $Q$ knots defined by:
\begin{equation*}
    \log(\lambda_{0k}(t)) = \exp\left(\sum_{q=1}^{Q+4} \eta_{qk} B_q(t,\nu_k)\right),
    \label{splines}
\end{equation*}
where $B_q(t,\nu_k)$ is the q-th basis function of B-splines with the knot vector $\nu_k$ and $\eta_{qk}$ is the associated parameter to be estimated.

\subsection{Estimation procedure}\label{subsec2}
\label{Proc_sec}
Let $\theta$ be the set of parameters to be estimated including parameters of the Cholesky decomposition of the covariance matrix of the random effects, $\beta$, $\mu$, $\alpha^\top = (\alpha_{11},\alpha_{21},\alpha_{\sigma 1},\alpha_{12},\alpha_{22},\alpha_{\sigma 2})$, $\gamma^\top = (\gamma_1,\gamma_2)$ and the parameters of the two baseline risk functions. Considering the frequentist approach, the parameter estimation is obtained by maximizing the likelihood function. The contribution of individual $i$ to the marginal likelihood is defined by:
\begin{eqnarray*}
\mathcal{L}_i(\theta;Y_i,T_i,\delta_i) & = & \int p(Y_i,T_i,\delta_i |b_i, \tau_i;\theta) f(b_i,\tau_i;\theta)db_id\tau_i \\
                      & = & \int f(Y_i|b_i,\tau_i;\theta)\exp\left(-\sum_{k=1}^2 \Lambda_{ik}(T_i|b_i,\tau_i;\theta)\right)\prod_{k=1}^{2}\lambda_{ik}(T_i|b_i,\tau_i;\theta)^{\mathbb{1}_{\delta_i=k}}f(b_i,\tau_i;\theta)db_id\tau_i,
\end{eqnarray*}

with $f(b_i,\tau_i;\theta)$ a multivariate Gaussian density and $f(Y_i|b_i,\tau_i;\theta)=\prod_{j=1}^{n_i}f(Y_{ij}|b_i,\tau_i;\theta)$ where $f(Y_{ij}|b_i,\tau_i;\theta)$ is a univariate Gaussian density. For $k \in \{1,2\}$, $\Lambda_{ik}(T_i|b_i,\tau_i;\theta)$ is the cumulative risk function given by:
\begin{equation}
\label{eq:cumfunct}
    \Lambda_{ik}(t|b_i,\tau_i;\theta) = \int_0^t\lambda_{ik}(u|b_i,\tau_i;\theta)du
\end{equation}

In cohort studies, data are frequently left-truncated. Left-truncation arises as soon as the time scale is not the time since inclusion and the subjects are enrolled only if they are free of the event at inclusion \citep{betensky_2015}. This is the case in most studies where the time-scale is age. To deal with left-truncation (also called delayed entry), the individual contribution to the likelihood must be divided by the probability to be free of any event at entry time $T_{0i}$:
\color{black}

\begin{equation*}
    \mathcal{L}_i^{DE}(\theta;Y_i,T_i,\delta_i) = \frac{\mathcal{L}_i(\theta;Y_i,T_i,\delta_i)}{\int \exp(-\Lambda_{i1}(T_{0i}|b_i,\tau_i;\theta)-\Lambda_{i2}(T_{0i}|b_i,\tau_i;\theta))f(b_i,\tau_i;\theta)db_id\tau_i}
\end{equation*}

Because the integral on the random effects does not have an analytical solution, the integral is computed by a Quasi Monte Carlo (QMC) approximation \citep{pan_quasi-monte_2007}, using deterministic quasi-random sequences. The approximation of the integral is defined by:
\begin{equation*}
   \mathcal{L}_i(\theta;Y_i,T_i,\delta_i) \simeq \frac{1}{S} \sum_{s = 1}^Sp(Y_i,T_i,\delta_i|b_i^s,\tau_i^s;\theta)
\end{equation*} 
with $(b_i^1,...,b_i^S)$ and $(\tau_i^1,...,\tau_i^S)$ are draws of a S-sample in the sobol sequel for the distribution $f(b_i,\tau_i;\theta)$.\\

To approximate the cumulative risk function given in equation \eqref{eq:cumfunct}, we use the Gauss-Kronrod quadrature approximation with 15 points \citep{gonnet_2012}.

Parameter estimation is obtained by maximizing the log-likelihood function $\ell(\theta;Y_i,T_i,\delta_i) = \log\left( \prod_{i=1}^N 
\mathcal{L}_i(\theta;Y_i,T_i,\delta_i)\right)$. The maximization is performed using the \texttt{marqLevAlg} R-package based on the Marquardt-Levenberg algorithm \citep{philipps_2021}. The latter is a robust variant of the Newton-Raphson algorithm \citep{levenberg_1944, marquardt_1963} which iteratively updates the parameters $\theta$ to be estimated until convergence with the following formula at iteration $l+1$:
\begin{equation*}
    \theta^{(l+1)} = \theta^{(l)} - \psi_l(\Tilde{H}(\theta^{(l)}))^{-1}\nabla(\ell(\theta^{(l)}))
\end{equation*}
where $\theta^{(l)}$ is the set of parameters at iteration $l$, $\nabla\left(\ell\left(\theta^{(l)}\right)\right)$ the gradient of the log-likelihood at iteration $l$ and $\widetilde{H}\left(\theta^{(l)}\right)$ the inflated Hessian matrix where the diagonal terms of the Hessian matrix $H\left(\theta^{(l)}\right)$  are replaced by :
\begin{equation*}
    \widetilde{H}\left(\theta^{(l)}\right)_{i i}=H\left(\theta^{(l)}\right)_{i i}+\phi_{l}\left[\left(1-\rho_{l}\right)\left|H\left(\theta^{(l)}\right)_{i i}\right|+\rho_{l} \operatorname{tr}\left(\theta^{(l)}\right)\right]. 
\end{equation*}
The scalars $\psi_l$, $\phi_l$ and $\rho_l$ are internally determined at each iteration $l$ to ensure that $\widetilde{H}\left(\theta^{(l)}\right)$ be definite-positive, $\widetilde{H}\left(\theta^{(l)}\right)$ approaches $H\left(\theta^{(l)}\right)$ when $\theta$ approaches $\widehat{\theta}$ and insure improvement of the likelihood at each iteration. Stringent convergence criteria are used, relying on parameter and function stability, and the relative distance to the maximum computed from the first and second derivatives of the log-likelihood which must not exceed a threshold $\varepsilon_d$: $\frac{\nabla\left(\ell\left(\theta^{(l)}\right)\right)\left(H\left(\theta^{(l)}\right)\right)^{-1} \nabla\left(\ell\left(\theta^{(l)}\right)\right)}{m}<\varepsilon_d$, with $m$ the number of parameters. This algorithm was previously compared to other algorithms (EM, BFGS and L-BFGS-B) and the results showed that this algorithm was the most reliable \citep{philipps_2021}.\\

The variances of the estimates are estimated by computing the inverse of the Hessian matrix computed by finite differences. The variances of the estimated parameters from the covariance matrix of the random effects are computed using the Delta-Method \citep{meyer_2013}.\\

To limit computation time and insure precise estimates of parameters and their standard error, we propose a two-step estimation procedure. In the first step, we applied the Marquardt-algorithm with a small number $S1$ of QMC draws (e.g. $S1 = 500$) until convergence is achieved. In the second step, we consider a number $S2>S1$ of QMC draws (e.g. $S2 = 5000$) to improve the accuracy of the computation of the standard error of the estimates that requires numerical derivation. If the Hessian matrix is not invertible, a few additional iterations are performed until invertibility is achieved. The selection of the number of QMC draws $S1$ and $S2$ significantly impacts computation time. Therefore, for a model selection step, we recommend that users compare various models based on results obtained in step 1 (using likelihood or information criteria) with a small value for $S1$. It is advisable to perform step 2, which involves a larger number of QMC draws, exclusively for the final model selected. 
\\

We implemented the computation of individual probability of having event $k$ between time $s$ and $s+t$ given that the subject $i$ did not experience any event before time $s$, its trajectory of marker until time $s$, $\mathcal{Y}_i(s)$, and the set of estimated parameters. The prediction is defined for subject $i$ by:

\begin{align}
   \pi_i(s,t;\widehat{\theta})&= P(s<T_i<s+t, \delta_i = k | T_i >s, \mathcal{Y}_i(s), \widehat{\theta}) \nonumber \\
   &= \frac{\int \left[\int_s^{s+t} \exp{(-\sum_{c=1}^2 \Lambda_{ic}(u|b_i,\tau_i,\widehat{\theta}))}\lambda_{ik}(u|b_i,\tau_i,\widehat{\theta})du \right]f(\mathcal{Y}_i(s)|b_i,\tau_i,\widehat{\theta})f(b_i,\tau_i|\widehat{\theta})db_id\tau_i}{\int\exp{(-\sum_{c=1}^2 \Lambda_{ic}(s|b_i,\tau_i,\widehat{\theta}))}f(\mathcal{Y}_i(s)|b_i,\tau_i,\widehat{\theta})f(b_i,\tau_i|\widehat{\theta})db_id\tau_i}
   \label{pred_st}
\end{align}

As previously, the integral over the random effect is computed by QMC approximation and the integral over time with the Gauss-Kronrod quadrature.\\

The corresponding 95\% confidence interval of predictions is obtained by the following Monte Carlo algorithm.\\
For $L$ large enough and $l  = 1,...,L$ ($L$ = 1000 for instance): 
\begin{itemize}
    \item Generate $\widetilde{\theta}^{(l)} \sim \mathcal{N}(\widehat{\theta},V(\widehat{\theta}))$ where $V(\widehat{\theta})$ is given by the inverse of the Hessian matrix at $\widehat{\theta}$
    \item Compute $\widetilde{\pi}^{(l)}_i(s,t;\widetilde{\theta}^{(l)})$ from equation \eqref{pred_st}
    \item Compute the 95\% confidence interval from the 2.5th and 97.5th percentiles of the L-sample of $\widetilde{\pi}^{(l)}_i(s,t;\widetilde{\theta}^{(l)})$\\
\end{itemize}

\subsection{Software}
The R-package \texttt{FlexVarJM} has been developed for the estimation of the model, the prediction of the subject-specific random effects, and the computation of the individual predicted probabilities of events. The package allows estimation of a model with an unconstrained time-trend for the marker trajectory, one or two events with exponential, Weibull or B-splines baseline risk functions, and a flexible dependence structure between the events and the marker (possibly including the current value, the current slope and the subject-specific time-dependent variability). The development version of \texttt{FlexVarJM} is available on Github at the following link: \url{https://github.com/LeonieCourcoul/FlexVarJM} and the fixed version can be installed from CRAN \citep{courcoul_2023}.

\section{Simulations}
\label{SimuPart}
 
In order to evaluate the performance of the estimation procedure, we performed a simulation study using a design similar to the application data.

\subsection{Design of simulations}

\noindent Visit times were generated using a uniform distribution centered around each specified time, with a variation of approximately one month in either direction. For each visit time, one measurement of the marker was generated, using a linear mixed-effects model with fixed and random intercept and slope, and heterogeneous variance:
\begin{equation}
\label{simumod1}
\left\{
    \begin{array}{ll}
         Y(t_{ij}) = \beta_0 + b_{0i} + (\beta_1 + b_{1i})\times t_{ij} + \epsilon_{i}(t_{ij})\\
        \epsilon_{i}(t_{ij}) \sim \mathcal{N}(0,\sigma_i^2(t_{ij})) \hspace{3mm} \text{with} \hspace{3mm} \log(\sigma_i(t_{ij}))  = \mu_0 + \tau_{0i} + (\mu_1 + \tau_{1i})\times t_{ij}
    \end{array}
\right.  
\end{equation}

with $b_i=\left(b_{0i},b_{1i}\right)^{\top}$ and $\tau_i=\left(\tau_{0i},\tau_{1i}\right)^{\top}$.
%
Competing event times $T_{ik}^*$ $(k = 1,2)$ were generated using the  Brent’s univariate root-finding method \citep{brent_1973} according to the following proportional hazards models:

\begin{equation}
\label{simumod2}
    \lambda_{ik}(t)= \lambda_{0k}(t) \exp(\alpha_{1k}\tilde{y}_i(t) +  \alpha_{2k}\tilde{y}'_i(t) + \alpha_{\sigma k} \sigma_i(t) )
\end{equation}

with $\lambda_{0k}(t) = \kappa_k t^{\kappa_k-1}e^{\zeta_{0k}}$ being a Weibull function. Individuals were censored at $C_i$ the last visit observed in the dataset. Finally, the observed time was defined by $T_i = \min(T_{i1}^*,T_{i2}^*,C_i)$. Measures of the marker $Y$ posterior to $T_i$ were removed from the datasets.\\

Five scenarii were considered varying the number of repeated measures and the correlation structure between the random effects: 
\begin{itemize}
    \item Scenario A: a maximum of 7 times of measurement, at 0-year, 0.5 year and then one per year until 5 years; with random effects $b_i$ independent of $\tau_i$.
    \item Scenario B: a maximum of 13 times of measurements, at 0 year, every 3 months the first year and then twice per year until 5 years (mimicking PROGRESS); with random effects $b_i$ independent of $\tau_i$.
    \item Scenario C: same time points as in Scenario A; with correlated random effects.
    \item Scenario D: same time points as in Scenario B; with correlated random effects.
    \item Scenario E (Misspecified model): marker generated with a quadratic trend but estimated with a linear trend; one event; same time points as in Scenario B; with random effects $b_i$ independent of $\tau_i$.
\end{itemize}

\noindent For each scenario, 300 datasets of 500 and 1000 subjects were generated. Parameter values for data generation are indicated in the tables of results. The models were estimated with the estimation procedure presented in Section \ref{Proc_sec}, given $S1 = 500$ and $S2=5000$ draws for the QMC integration approximation.

\subsection{Results}

Tables \ref{t:SA500}, \ref{t:SB500} and \ref{t:SC500} report the mean estimates, the empirical and mean asymptotic standard error of the estimated parameters and the coverage rate of their 95\% confidence intervals for scenario A, B and C on 500 individuals. Results for scenario D and for larger sample size are in the Supporting Information (tables S1 to S5). The estimation procedure provided satisfactory results for the four sets of simulations. Indeed, the bias was minimal, the mean asymptotic and the empirical standard errors were close, and the coverage rates of the 95\% confidence interval were close to the nominal value. We only observed slight under coverage of the confidence interval for some parameters in the covariance matrix of the random effects that tend to reduce for larger sample size (N=1000 subjects, see Tables S1, S2 and S3 in Supporting Information). We can note that the bias is minimal from the first step but the second step helps to reduce the difference between the mean asymptotic and the empirical standard deviations and thus improve the coverage rates. This simulation study also illustrates the impact of the choice of $S1$ and $S2$ on the computation time: for scenario A, the medians of computation time are around 13 minutes (25 iterations in median) and 9 minutes (1 iteration in median) respectively for step 1 and 2. Finally, Scenario E was performed to evaluate the impact of a misspecified marker trajectory (quadratic versus linear time trend). As expected, the estimates of the fixed effects and covariance matrix for $b_i$ in the mixed models are biased, but the estimates of the model for the residual variance and of the time-to-event model are robust (Table S6 in Supporting Information).\\
\color{black}

\begin{table}[]
\caption{Simulation results for scenario A with 500 subjects (7 measures, $b_i$ and $\tau_i$ independent).*}
\label{t:SA500}
\scalebox{0.7}{\begin{tabular}{|lll|cccc|cccc|}
\hline
\multicolumn{3}{|c|}{\textbf{Parameter}}                                                                                         & \multicolumn{4}{|c|}{\textbf{Step 1}}                                                & \multicolumn{4}{|c|}{\textbf{Step 2}}                                                \\ \hline
\multicolumn{2}{|l}{\textit{\textbf{}}}                                                                        & \textbf{True}  & \textbf{Mean} & \textbf{Empirical} & \textbf{Mean asymptotic} & \textbf{Coverage}  & \textbf{Mean} & \textbf{Empirical} & \textbf{Mean asymptotic} & \textbf{Coverage}  \\
\multicolumn{2}{|l}{\textit{\textbf{}}}                                                                        & \textbf{value} & \textbf{}     & \textbf{SE}        & \textbf{SE}              & \textbf{rate ($\%$)} & \textbf{}     & \textbf{SE}        & \textbf{SE}              & \textbf{rate ($\%$)} \\ \hline
\multicolumn{11}{|l|}{Longitudinal submodel}                                                                                                                                                                                                                                                               \\ \hline
\textit{Intercept}                  & $\beta_0$                                                 & 142            & 142.1       & 0.736              & 0.717                    & 94.31             & 142.1      & 0.730              & 0.728                    & 94.98             \\
\textit{Slope}                      & $\beta_1$                                                & 3              & 2.943         & 0.288              & 0.253                    & 89.97             & 2.945         & 0.280              & 0.271                    & 92.64             \\
\textit{Variability}                & $\mu_0$                                                   & 2.4            & 2.396         & 0.027              & 0.026                    & 94.65             & 2.395         & 0.027              & 0.027                    & 95.32             \\
\textit{}                           & $\mu_1$                                                  & 0.05           & 0.050         & 0.017              & 0.015                    & 92.98             & 0.051         & 0.016              & 0.016                    & 94.31             \\
$\Sigma_b$   & $\sigma_{b_0}^2$                      & 207.36         & 205.0       & 17.31             & 16.17                   & 91.97             & 205.1       & 17.19             & 16.60                   & 91.97             \\
\textit{}                           & $\sigma_{b_0b_1}$                                    & -17.28         & -16.06       & 4.132              & 3.663                    & 87.96             & -16.04       & 3.936              & 4.020                    & 92.64             \\
\textit{}                           & $\sigma_{b_1}^2  $                    & 9.224         & 9.351        & 1.673             & 1.308                   & 84.95             & 9.279        & 1.536             & 1.465                   & 91.64             \\

$\Sigma_{\tau}$ & $\sigma_{\tau_0}^2$    & 0.0001         & 0.004        & 0.006             & 0.006                   & 98.66             & 0.004        & 0.007             & 0.006                   & 98.33             \\
\textit{}                           & $\sigma_{\tau_0\tau_1}$ & -0.0006        & -0.002       & 0.005             & 0.005                   & 92.64             & -0.003       & 0.005             & 0.006                   & 95.99             \\
\textit{}                           & $\sigma_{\tau_1}^2$    & 0.0157         & 0.016        & 0.005             & 0.004                   & 91.97             & 0.017        & 0.005             & 0.005                   & 93.31             \\ \hline
\multicolumn{11}{|l|}{\textit{Survival submodel 1}}                                                                                                                                                                                                                                                        \\ \hline
\textit{Current variance}           & $\alpha_{\sigma1}$                       & 0.07           & 0.064         & 0.041              & 0.039                    & 94.65             & 0.064         & 0.041              & 0.039                    & 95.32             \\
\textit{Current value}              & $\alpha_{11} $                                          & 0.02           & 0.020         & 0.008              & 0.007                    & 95.65             & 0.020         & 0.008              & 0.007                    & 95.32             \\
\textit{Current slope}              & $\alpha_{21}  $                                         & 0.01           & 0.008         & 0.072              & 0.066                    & 94.31             & 0.007         & 0.071              & 0.067                    & 95.99             \\
\textit{Weibull}                    & $\sqrt{\kappa_1}  $                      & 1.1            & 1.099         & 0.056              & 0.059                    & 97.66             & 1.098         & 0.056              & 0.059                    & 97.99             \\
\textit{}                           & $\zeta_{01} $                                           & -7             & -6.884        & 1.257              & 1.199                    & 94.31             & -6.885        & 1.257              & 1.204                    & 94.65             \\ \hline
\multicolumn{11}{|l|}{\textit{Survival submodel 2}}                                                                                                                                                                                                                                                        \\ \hline
\textit{Current variance}           & $\alpha_{\sigma2}$                   & 0.15           & 0.169         & 0.091              & 0.046                    & 92.31             & 1.678         & 0.091              & 0.054                    & 96.32             \\
\textit{Current value}              & $\alpha_{12}$                                      & -0.01          & -0.012        & 0.014              & 0.009                    & 96.32             & -0.012        & 0.015              & 0.010                    & 96.66             \\
\textit{Current slope}              & $\alpha_{22}$                                        & -0.14          & -0.171        & 0.173              & 0.086                    & 92.64             & -0.167        & 0.170              & 0.095                    & 94.65             \\
\textit{Weibull}                    & $\sqrt{\kappa_2}$                      & 1.3            & 1.310         & 0.102              & 0.079                    & 96.99             & 1.311         & 0.105              & 0.084                    & 97.99             \\
\textit{}                           & $\zeta_{02}$                                         & -4             & -4.075        & 1.389              & 1.366                    & 95.99             & -4.075        & 1.388              & 1.401                    & 95.65             \\ \hline

\end{tabular}}

\vspace{3mm}
SE: Standard Error; Coverage rate: coverage rate of the $95\%$ confidence interval.\\
* Results for 299 replicates with complete convergence over 300.
\end{table}


\begin{table}[]
\caption{Simulation results for scenario B with 500 subjects (13 measures, $b_i$ and $\tau_i$ independent).*}
\label{t:SB500}
\scalebox{0.7}{\begin{tabular}{|lll|cccc|cccc|}
\hline
\multicolumn{3}{|c|}{\textbf{Parameter}}                                                                                         & \multicolumn{4}{c}{\textbf{Step 1}}                                                & \multicolumn{4}{|c|}{\textbf{Step 2}}                                                \\ \hline
\multicolumn{2}{|l}{\textit{\textbf{}}}                                                                        & \textbf{True}  & \textbf{Mean} & \textbf{Empirical} & \textbf{Mean asymptotic} & \textbf{Coverage}  & \textbf{Mean} & \textbf{Empirical} & \textbf{Mean asymptotic} & \textbf{Coverage}  \\
\multicolumn{2}{|l}{\textit{\textbf{}}}                                                                        & \textbf{value} & \textbf{}     & \textbf{SE}        & \textbf{SE}              & \textbf{rate ($\%$)} & \textbf{}     & \textbf{SE}        & \textbf{SE}              & \textbf{rate ($\%$)} \\ \hline
\multicolumn{11}{|l|}{Longitudinal submodel}                                                                                                                                                                                                                                                               \\ \hline
\textit{Intercept}                  & $\beta_0$                                             & 142            & 142.0       & 0.779              & 0.655                    & 90.67             & 142.0       & 0.767              & 0.721                    & 92.67             \\
\textit{Slope}                      & $\beta_1$                                                 & 3              & 2.996         & 0.252              & 0.201                    & 90.00               & 3.000         & 0.248              & 0.235                    & 93.33           \\
\textit{Variability}                & $\mu_0$                                                   & 2.4            & 2.402         & 0.019              & 0.018                    & 92.00               & 2.401         & 0.019              & 0.018                    & 92.67             \\
\textit{}                           & $\mu_1$                                                   & 0.05           & 0.047         & 0.012              & 0.011                    & 92.00               & 0.049         & 0.012              & 0.012                    & 94.00               \\
$\Sigma_b$   & $\sigma_{b_0}^2$                      & 207.36         & 208.1       & 17.79             & 14.18                  & 87.00               & 208.0       & 17.29             & 15.92                   & 92.33             \\
\textit{}                           & $\sigma_{b_0b_1}$                                     & -17.28         & -15.74       & 4.165              & 2.954                    & 77.00               & -15.85       & 4.022              & 3.614                    & 85.00               \\
\textit{}                           & $\sigma_{b_1}^2   $                   & 9.224         & 9.236        & 1.476             & 0.948                   & 79.00               & 9.256        & 1.322             & 1.246                   & 89.67             \\
$\Sigma_{\tau}$ & $\sigma_{\tau_0}^2$    & 0.0001         & 0.002        & 0.003             & 0.002                   & 97.33             & 0.002        & 0.003             & 0.003                   & 97.67             \\
\textit{}                           & $\sigma_{\tau_{0}\tau_{1}}$ & -0.0006        & -0.001       & 0.003             & 0.003                   & 91.33             & -0.001       & 0.003             & 0.004                   & 93.67             \\
\textit{}                           & $\sigma_{\tau_1}^2$    & 0.0157         & 0.015        & 0.003             & 0.003                   & 88.67             & 0.016        & 0.003             & 0.003                   & 94.67             \\ \hline
\multicolumn{11}{|l|}{\textit{Survival submodel 1}}                                                                                                                                                                                                                                                        \\ \hline
\textit{Current variance}           & $\alpha_{\sigma1}$                       & 0.07           & 0.063         & 0.030              & 0.028                    & 92.67             & 0.065         & 0.029              & 0.028                    & 93.67             \\
\textit{Current value}              & $\alpha_{11}   $                                        & 0.02           & 0.020         & 0.006              & 0.007                    & 96.33             & 0.020         & 0.006              & 0.007                    & 96.33             \\
\textit{Current slope}              & $\alpha_{21}    $                                       & 0.01           & 0.007         & 0.057              & 0.053                    & 96.33             & 0.007         & 0.055              & 0.055                    & 96.00               \\
\textit{Weibull}                    & $\sqrt{\kappa_1}  $                      & 1.1            & 1.106         & 0.055              & 0.054                    & 94.00               & 1.105         & 0.055              & 0.055                    & 95.00               \\
\textit{}                           & $\zeta_{01} $                                           & -7             & -6.912        & 0.992              & 1.042                    & 95.67             & -6.914        & 0.992              & 1.050                    & 95.67             \\ \hline
\multicolumn{11}{|l|}{\textit{Survival submodel 2}}                                                                                                                                                                                                                                                        \\ \hline
\textit{Current variance}           & $\alpha_{\sigma2}    $                   & 0.15           & 0.158         & 0.034              & 0.030                    & 92.67             & 0.159         & 0.031              & 0.032                    & 97.33             \\
\textit{Current value}              & $\alpha_{12}        $                                   & -0.01          & -0.010        & 0.008              & 0.008                    & 94.67             & -0.010        & 0.008              & 0.008                    & 95.33             \\
\textit{Current slope}              & $\alpha_{22} $                                          & -0.14          & -0.146        & 0.064              & 0.061                    & 94.00               & -0.145        & 0.063              & 0.064                    & 95.00               \\
\textit{Weibull}                    & $\sqrt{\kappa_2}    $                    & 1.3            & 1.299         & 0.067              & 0.067                    & 95.67             & 1.299         & 0.066              & 0.068                    & 96.33             \\
\textit{}                           & $\zeta_{02}    $                                        & -4             & -4.104        & 1.173              & 1.111                    & 93.67             & -4.107        & 1.173              & 1.139                    & 93.67             \\ \hline

\end{tabular}}

\vspace{3mm}
SE: Standard Error; Coverage rate: coverage rate of the $95\%$ confidence interval.\\
* Results for 300 replicates with complete convergence over 300.
\end{table}


\begin{table}[]
\caption{Simulation results for scenario C with 500 subjects (7 measures, $b_i$ and $\tau_i$ correlated).*}
\label{t:SC500}
\scalebox{0.7}{\begin{tabular}{|lllcccccccc|}
\hline
\multicolumn{3}{|c|}{\textbf{Parameter}}                                                                                                   & \multicolumn{4}{c|}{\textbf{Step 1}}                                                                                                                                & \multicolumn{4}{c|}{\textbf{Step 2}}                                                                                                                                \\ \hline
\multicolumn{2}{|l}{\textit{\textbf{}}}                                                              & \multicolumn{1}{l|}{\textbf{True}}  & \multicolumn{1}{l}{\textbf{Mean}} & \multicolumn{1}{l}{\textbf{Empirical}} & \multicolumn{1}{l}{\textbf{Mean asymptotic}} & \multicolumn{1}{l|}{\textbf{Coverage}}  & \multicolumn{1}{l}{\textbf{Mean}} & \multicolumn{1}{l}{\textbf{Empirical}} & \multicolumn{1}{l}{\textbf{Mean asymptotic}} & \multicolumn{1}{l|}{\textbf{Coverage}}  \\
\multicolumn{2}{|l}{\textit{\textbf{}}}                                                              & \multicolumn{1}{l|}{\textbf{value}} & \multicolumn{1}{l}{\textbf{}}     & \multicolumn{1}{l}{\textbf{SE}}        & \multicolumn{1}{l}{\textbf{SE}}              & \multicolumn{1}{l|}{\textbf{rate ($\%$)}} & \multicolumn{1}{l}{\textbf{}}     & \multicolumn{1}{l}{\textbf{SE}}        & \multicolumn{1}{l}{\textbf{SE}}              & \multicolumn{1}{l|}{\textbf{rate ($\%$)}} \\ \hline
\multicolumn{11}{|l|}{Longitudinal submodel}                                                                                                                                                                                                                                                                                                                                                                                                                                           \\ \hline
\textit{Intercept}         & $\beta_0 $                                                & \multicolumn{1}{l|}{142}            & 141.9                           & 0.833                                  & 0.742                                        & \multicolumn{1}{c|}{92.28}             & 141.9                           & 0.820                                  & 0.756                                        & 93.33                                  \\
\textit{Slope}             & $\beta_1   $                                              & \multicolumn{1}{l|}{3}              & 3.019                             & 0.320                                  & 0.282                                        & \multicolumn{1}{c|}{90.61}             & 3.023                             & 0.314                                  & 0.290                                        & 92.00                                    \\
\textit{Variability}       & $\mu_0 $                                                  & \multicolumn{1}{l|}{2.4}            & 2.401                             & 0.035                                  & 0.033                                        & \multicolumn{1}{c|}{93.62}             & 2.399                             & 0.035                                  & 0.033                                        & 94.33                                  \\
\textit{}                  & $\mu_1$                                                   & \multicolumn{1}{l|}{0.05}           & 0.050                             & 0.016                                  & 0.015                                        & \multicolumn{1}{c|}{92.62}             & 0.050                             & 0.016                                  & 0.016                                        & 93.67                                  \\
$\Sigma_{b\tau}$   & $\sigma_{b_0}^2$                      & \multicolumn{1}{l|}{210.25}         & 209.7                          & 20.69                                 & 16.90                                       & \multicolumn{1}{c|}{88.23}             & 209.4                           & 20.39                                 & 17.72                                       & 91.33                                  \\
\textit{}                  & $\sigma_{b_0b_1}$                                     & \multicolumn{1}{l|}{-15.95}         & -15.43                           & 4.734                                  & 3.942                                        & \multicolumn{1}{c|}{88.59}             & -15.48                           & 4.398                                  & 4.209                                        & 92.33                                  \\
                           & $\sigma_{b_0\tau_0}     $              & \multicolumn{1}{l|}{2.9}            & \multicolumn{1}{c}{2.796}         & \multicolumn{1}{c}{0.610}              & \multicolumn{1}{c}{0.521}                    & \multicolumn{1}{c|}{88.93}             & \multicolumn{1}{c}{2.812}         & \multicolumn{1}{c}{0.592}              & \multicolumn{1}{c}{0.533}                    & \multicolumn{1}{c|}{89.67}             \\
                           & $\sigma_{b_0\tau_1} $                  & \multicolumn{1}{l|}{-0.145}         & \multicolumn{1}{c}{-0.129}       & \multicolumn{1}{c}{0.251}             & \multicolumn{1}{c}{0.210}                   & \multicolumn{1}{c|}{89.54}             & \multicolumn{1}{c}{-0.115}       & \multicolumn{1}{c}{0.240}             & \multicolumn{1}{c}{0.220}                   & \multicolumn{1}{c|}{93.00}               \\
\textit{}                  &$ \sigma_{b_1}^2    $                  & \multicolumn{1}{l|}{9.05}           & 9.181                             & 1.745                                  & 1.371                                        & \multicolumn{1}{c|}{85.91}             & 9.084                             & 1.599                                  & 1.479                                        & 91.33                                  \\
                           & $\sigma_{b_1\tau_0}       $            & \multicolumn{1}{l|}{-0.304}         & \multicolumn{1}{c}{-0.295}       & \multicolumn{1}{c}{0.178}             & \multicolumn{1}{c}{0.157}                   & \multicolumn{1}{c|}{91.95}             & \multicolumn{1}{c}{-0.291}       & \multicolumn{1}{c}{0.169}             & \multicolumn{1}{c}{0.162}                   & \multicolumn{1}{c|}{94.67}             \\
                           & $\sigma_{b_1\tau_1}  $                 & \multicolumn{1}{l|}{0.067}          & \multicolumn{1}{c}{0.069}        & \multicolumn{1}{c}{0.073}             & \multicolumn{1}{c}{0.061}                   & \multicolumn{1}{c|}{88.93}             & \multicolumn{1}{c}{0.065}        & \multicolumn{1}{c}{0.068}             & \multicolumn{1}{c}{0.063}                   & \multicolumn{1}{c|}{92.00}               \\
 			   & $\sigma_{\tau_0}^2 $   & \multicolumn{1}{l|}{0.1309}         & 0.123                            & 0.029                                 & 0.026                                       & \multicolumn{1}{c|}{87.53}             & 0.128                            & 0.029                                 & 0.028                                       & 91.67                                  \\
\textit{}                  & $\sigma_{\tau_0\tau_1}$ & \multicolumn{1}{l|}{-0.0206}        & -0.019                           & 0.011                                 & 0.008                                       & \multicolumn{1}{c|}{87.53}             & -0.021                           & 0.010                                 & 0.010                                       & 94.00                                    \\
\textit{}                  & $\sigma_{\tau_1}^2$    & \multicolumn{1}{l|}{0.0141}         & 0.014                            & 0.005                                 & 0.004                                       & \multicolumn{1}{c|}{80.87}             & 0.015                            & 0.005                                 & 0.005                                       & 94.33                                  \\ \hline
\multicolumn{11}{|l|}{\textit{Survival submodel 1}}                                                                                                                                                                                                                                                                                                                                                                                                                                    \\ \hline
\textit{Current variance}  & $\alpha_{\sigma1} $                      & \multicolumn{1}{l|}{0.07}           & 0.065                             & 0.051                                  & 0.046                                        & \multicolumn{1}{c|}{93.29}             & 0.067                             & 0.050                                  & 0.047                                        & 94.67                                  \\
\textit{Current value}     & $\alpha_{11}$                                           & \multicolumn{1}{l|}{0.02}           & 0.021                             & 0.012                                  & 0.011                                        & \multicolumn{1}{c|}{93.29}             & 0.021                             & 0.012                                  & 0.011                                        & 93.33                                  \\
\textit{Current slope}     & $\alpha_{21}$                                           & \multicolumn{1}{l|}{0.01}           & -0.004                            & 0.077                                  & 0.076                                        & \multicolumn{1}{c|}{93.62}             & -0.051                            & 0.817                                  & 0.470                                        & 94.00                                    \\
\textit{Weibull}           & $\sqrt{\kappa_1}$                        & \multicolumn{1}{l|}{1.1}            & 1.094                             & 0.052                                  & 0.054                                        & \multicolumn{1}{c|}{96.31}             & 1.095                             & 0.052                                  & 0.055                                        & 96.67                                  \\
\textit{}                  & $\zeta_{01}$                                            & \multicolumn{1}{l|}{-7}             & -7.133                            & 1.437                                  & 1.296                                        & \multicolumn{1}{c|}{93.96}             & -6.958                            & 3.204                                  & 2.612                                        & 94.67                                  \\ \hline
\multicolumn{11}{|l|}{\textit{Survival submodel 2}}                                                                                                                                                                                                                                                                                                                                                                                                                                    \\ \hline
\textit{Current variance}  & $\alpha_{\sigma2}$                       & \multicolumn{1}{l|}{0.15}           & 0.166                             & 0.079                                  & 0.051                                        & \multicolumn{1}{c|}{91.28}             & 0.165                             & 0.077                                  & 0.057                                        & 95.00                                    \\
\textit{Current value}     &$ \alpha_{12}  $                                         & \multicolumn{1}{l|}{-0.01}          & -0.016                            & 0.018                                  & 0.013                                        & \multicolumn{1}{c|}{91.95}             & -0.012                            & 0.017                                  & 0.014                                        & 93.67                                  \\
\textit{Current slope}     &$ \alpha_{22}$                                           & \multicolumn{1}{l|}{-0.14}          & -0.154                            & 0.113                                  & 0.088                                        & \multicolumn{1}{c|}{95.64}             & -0.178                            & 0.486                                  & 0.253                                        & 95.67                                  \\
\textit{essai} & $\sqrt{\kappa_2}$ & \multicolumn{1}{l|}{1.3}            & 1.314                             & 0.078                                  & 0.069                                        & \multicolumn{1}{c|}{94.30}             & 1.314                             & 0.074                                  & 0.072                                        & 94.67                                  \\
\textit{}                  & $\zeta_{02}$                                            & \multicolumn{1}{l|}{-4}             & -4.084                            & 1.758                                  & 1.524                                        & \multicolumn{1}{c|}{94.30}             & -3.983                            & 2.469                                  & 2.109                                        & 95.00                                    \\ \hline

\hline
\end{tabular}}

\vspace{3mm}
SE: Standard Error; Coverage rate: coverage rate of the $95\%$ confidence interval.\\
* Results for 298 replicates with complete convergence over 300 for step 1 and for 300 replicates with complete convergence over 300 for step 2.
\end{table}

\section{Application}
\label{s:model}

\subsection{PROGRESS clinical trial}

We estimated the proposed model on the data from the PROGRESS clinical trial \citep{progress_2001} a blood-pressure lowering, multicentre, double-blind randomized placebo-controlled clinical trial including patients with a history of stroke or transient ischaemic attack within 5 years before inclusion. Patients were recruited between May 1995 and November 1997. The follow-up comprised five visits in the first year, then two visits each years until the end of the study or the occurrence of a major CVD (stroke, myocardial infarction and cerebral hemorrhage) or death. At each visit, blood pressure was measured twice and we analysed the mean of the two measurements at each time. Prior to randomization, eligible patients were subjected to a 4-week run-in phase to test their tolerance to the treatment. At randomization, patients assigned to the control group stopped the treatment. In order to avoid an effect of the change of therapy at randomization, we removed the blood pressure measure at randomization. Finally, the current study was conducted over 3710 Non Asian patients, 1856 for the controlled group and 1854 for the treatment group, and included 672 CVD and 150 deaths without CVD. There are 2525 (68\%) men and 1185 (32\%) women. The average age at entry in the study is 67 years old (sd = 9.8) with a minimum at 26 and a maximum at 91 years old.

\subsection{Specification of the model}
This study aimed to evaluate the impact of the blood pressure variability on the risk of CVD and death from other causes. To do so, we estimated the proposed joint model (Model CVCS+V, for current value, current slope and variance) with heterogeneous time-dependent variance defined by \eqref{Mixed} and \eqref{Surv} using the time since the first considered blood pressure measurement. The trajectory of blood pressure was described over time by a linear mixed effect model. The individual time trend of the marker and the variance were modelled by a linear trend. The baseline hazard functions of both events were defined by B-splines with three knots placed at the quantiles of the observed events. According to the AIC, the model with three knots was better than models with 1 or 5 knots for each baseline hazard function (respectively 298946.2, 298948.8 and 298951). The model allowed the risk of each event to depend on the time-dependent intra-subject variability, the individual current value and the current slope. The longitudinal submodel and the variance submodel were adjusted for treatment group and survival submodels were adjusted for treatment group, age at baseline and sex (male versus female):

\begin{equation*}
\left\{
    \begin{array}{ll}
         y(t_{ij}) = \beta_0 + b_{0i} + (\beta_1 + b_{1i})\times t_{ij} + \beta_2\times trt_i + \epsilon_{i}(t_{ij})\\
        \epsilon_{i}(t_{ij}) \sim \mathcal{N}(0,\sigma_i^2(t_{ij})) \hspace{3mm} \text{with} \hspace{3mm} \log(\sigma_i(t_{ij}))  = \mu_0 + \tau_{0i} + (\mu_1 + \tau_{1i})\times t_{ij} + \mu_2 \times trt_i\\
        \lambda_{ik}(t)= \lambda_{0k}(t) \exp(\gamma_{0k}trt_i + \gamma_{1k}male_i + \gamma_{2k}age + \alpha_{1k}\tilde{y}_i(t) +  \alpha_{2k}\tilde{y}'_i(t) + \alpha_{\sigma k} \sigma_i(t) )
    \end{array}
\right.  
\end{equation*}

The estimation was performed with $S1 = 500$ and $S2 = 10000$ draws of QMC to ensure a greater accuracy.\\

This model was compared to two classical joint model without heterogeneous variance , i.e. $\sigma_i^2(t_{ij})= \sigma^2$ for all $i=1,\ldots,N$ and $j=1,\ldots,n_i$. The first one allowed the risk of each event to depend only on the individual current value (Model CV) and the second one on both the individual current value and current slope (Model CVCS).\\

\subsection{Results}

The AIC from the complete model (298946.2) was clearly better than the AIC from the two joint models with a constant residual variance and either with a dependence on the current value only (302062.4) or with a dependance on the current value and the current slope (302011.6), showing the importance of taking into account a time-dependent subject-specific variance.\\

Table~\ref{App} provides estimates from the complete joint model and Table~\ref{DM} the covariance matrix of the random effects and their standard errors computed through the Delta-Method. Blood pressure was lower for individuals from the treatment group ($\beta_2 = -8.03$, $p-value <0.001$). The variance of the residual error was heterogeneous between the subjects ($\widehat{Var}(\tau_{0i}) = 0.13$, $sd = 7e-3$) and was lower for treated patients ($\hat{\mu_2} = -0.030$, $p-value = 0.028$). The risk of CVD events increased with age ($HR = 1.04$ for one year, $IC = [1.03;1.05]$) and was higher for men ($HR = 1.34$, $IC = [1.13;1.60]$). Adjusting for age, sex and treatment group, the risk of CVD was associated with the current blood pressure variance ($HR = 1.07$, $IC=[1.03;1.10]$): the higher the standard error of blood pressure, the higher the risk of CVD; and with the current slope of blood pressure ($HR = 0.86$, $IC=[0.82;0.90]$). This last result means that patients with a decreasing slope had a higher risk of CVD. These patients could be those with an history of hypertension. A similar effect is observed on the CVCS model (see Table S7 in Supporting Information). However, this risk did not depend on the current blood pressure value ($HR= 0.99$, $IC=[0.986;1.002]$). The risk of death from other causes was higher for older individuals ($HR = 1.05$, $IC=[1.03;1.07]$) and for men ($HR = 1.69$, $IC=[1.16;2.48]$). It was not associated with the treatment group. Moreover, it was associated with the current value of blood pressure: the  instantaneous risk of death was multiplied by 0.90 ($IC=[0.816;0.993]$) for each increase of 5 mmHg of the mean blood pressure. The risk of death was also associated with the current blood pressure variance (HR = 1.13, $IC=[1.05;1.21]$) and with the current slope ($HR = 0.89$, $IC=[0.798;0.998]$).\\

\begin{table}[h!]
\centering
\caption{Parameter estimates of the joint model on the Progress clinical trial data (CVCS+V model).}
\label{App}
\begin{tabular}{|lccc|}
\hline
\textbf{Parameter} & \textbf{Estimate} & \textbf{Standard error} & \textbf{p-value} \\
  \hline

\multicolumn{4}{|l|}{\textit{Survival submodel for CVD}} \\ 
\hspace{6mm} BP current variance &  0.064 & 0.017 & $<0.001$\\
\hspace{6mm} BP current value & -0.006 & 0.004 & $0.160$\\
\hspace{6mm} BP current slope &  -0.152 & 0.022 & $<0.001$\\
\hspace{6mm} treatment group &  -0.153 & 0.085 & $0.073$\\
\hspace{6mm} male & 0.296 & 0.089 & $<0.001$\\
\hspace{6mm} age & 0.038 & 0.005 & $<0.001$\\
\multicolumn{4}{|l|}{\textit{Survival submodel for Death}} \\
\hspace{6mm} BP current variance &  0.120 & 0.035 & $<0.001$\\
\hspace{6mm} BP current value & -0.021 & 0.010 & $0.030$\\
\hspace{6mm} BP current slope &  -0.114 & 0.057 & $0.045$\\
\hspace{6mm} treatment group &  -0.117 & 0.171 & $0.493$\\
\hspace{6mm} male & 0.527 & 0.194 & $0.006$\\
\hspace{6mm} age & 0.051 & 0.010 & $<0.001$\\
\multicolumn{4}{|l|}{\textit{Longitudinal submodel}} \\
\multicolumn{4}{|l|}{\underline{Blood Pressure Mean}} \\
\hspace{6mm} intercept & 142.5 & 0.330 & $<0.001$\\
\hspace{6mm} time & -0.104 & 0.072 & 0.150\\
\hspace{6mm} treatment group & -8.029 & 0.441 & $<0.001$\\
\multicolumn{4}{|l|}{\underline{Blood Pressure Residual Variance}} \\
\hspace{6mm} intercept & 2.341 & 0.012 & $<0.001$ \\
\hspace{6mm} time & 0.007 & 0.004 & 0.086 \\
\hspace{6mm} treatment group & -0.030 & 0.014 & 0.028 \\
\hline
\end{tabular}\\
\vspace{5mm}
BP: Blood Pressure
\end{table}

\begin{table}[h!]
\caption{Covariance matrix (and standard errors) of the random effects computed using the Delta-Method.}
    \centering
    \begin{align*}
    \Sigma & =  \left[\begin{array}{cccc}
Var(b_{0i}) &  &  &   \\
Cov(b_{0i},b_{1i}) & Var(b_{1i}) &  &  \\
Cov(b_{0i},\tau_{0i}) & Cov(b_{1i},\tau_{0i}) & Var(\tau_{0i}) &  \\
Cov(b_{0i},\tau_{1i}) & Cov(b_{1i},\tau_{1i}) & Cov(\tau_{0i},\tau_{1i}) & Var(\tau_{1i}) 
\end{array}\right] \\
& = \left[\begin{array}{cccc}
212.3_{(5.6)} &  &  &   \\
-19.0_{(1.2)} & 8.6_{(0.4)} &  &  \\
2.2_{(0.15)} & -0.29_{(0.04)} & 0.13_{(7e-3)} &  \\
-0.18_{(0.06)} & 0.16_{(0.02)} & -0.02_{(3e-3)} & 0.012_{(1e-3)} 
\end{array}\right]
\end{align*}
\label{DM}
\end{table}

\subsection{Goodness-of-fit assessment}
\label{s:fit}
To assess the fit of the time-to-event submodels, we computed for each event, the predicted cumulative hazard function at each event time by plugging the empirical Bayes estimates of the random effects in the formula for the risk function. Then we compared the mean of this predicted cumulative hazard function with its Nelson-Aalen estimator for the whole sample (Figure S1 of the Supporting Information) and stratified according to sex and randomization group (Figure S2). These figures show that the joint model adequately fitted both risks and that the proportional risk assumption was valid for each categorical variable.\\

To highlight the impact of adding a subject-specific and time-dependent residual variance in the mixed model, we computed the individual predictions of the marker over time for some selected subjects. The predicted value of blood pressure corresponds to the conditional expectation given the random effects, defined by $\hat{\mathbb{E}}(Y_i(t)|\widetilde{b}_i,\widetilde{\tau}_i)$ and the prediction interval around this predicted values is given by $\hat{\mathbb{E}}(Y_i(t)|\widetilde{b}_i,\widetilde{\tau}_i) \pm 1.96\hat{\mathbb{V}}(Y_i(t)|\widetilde{b}_i,\widetilde{\tau}_i)$. For each subject the empirical Bayes estimates of the random effects, denoted by $(\widetilde{b}_i, \widetilde{\tau}_i) = \text{argmax}f(b_i,\tau_i|Y_i,T_i,\delta_i) $, corresponds to the mode of their estimated conditional posterior given the data. They are computed by maximising $f(Y_i,T_i,\delta_i)f(b_i,\tau_i)$ with the Marquardt-Levenberg algorithm.\\

For some selected subjects still at risk at 3 years, Figure \ref{s:fit} presents the predicted values and their confidence intervals from the models with and without subject-specific residual variance (CVCS+V and CVCS). It shows that assuming a time-dependent and subject-specific residual variability allows a better fit of the uncertainty around the individual prediction.\\

\begin{figure}[!ht]
    \centering
    \includegraphics{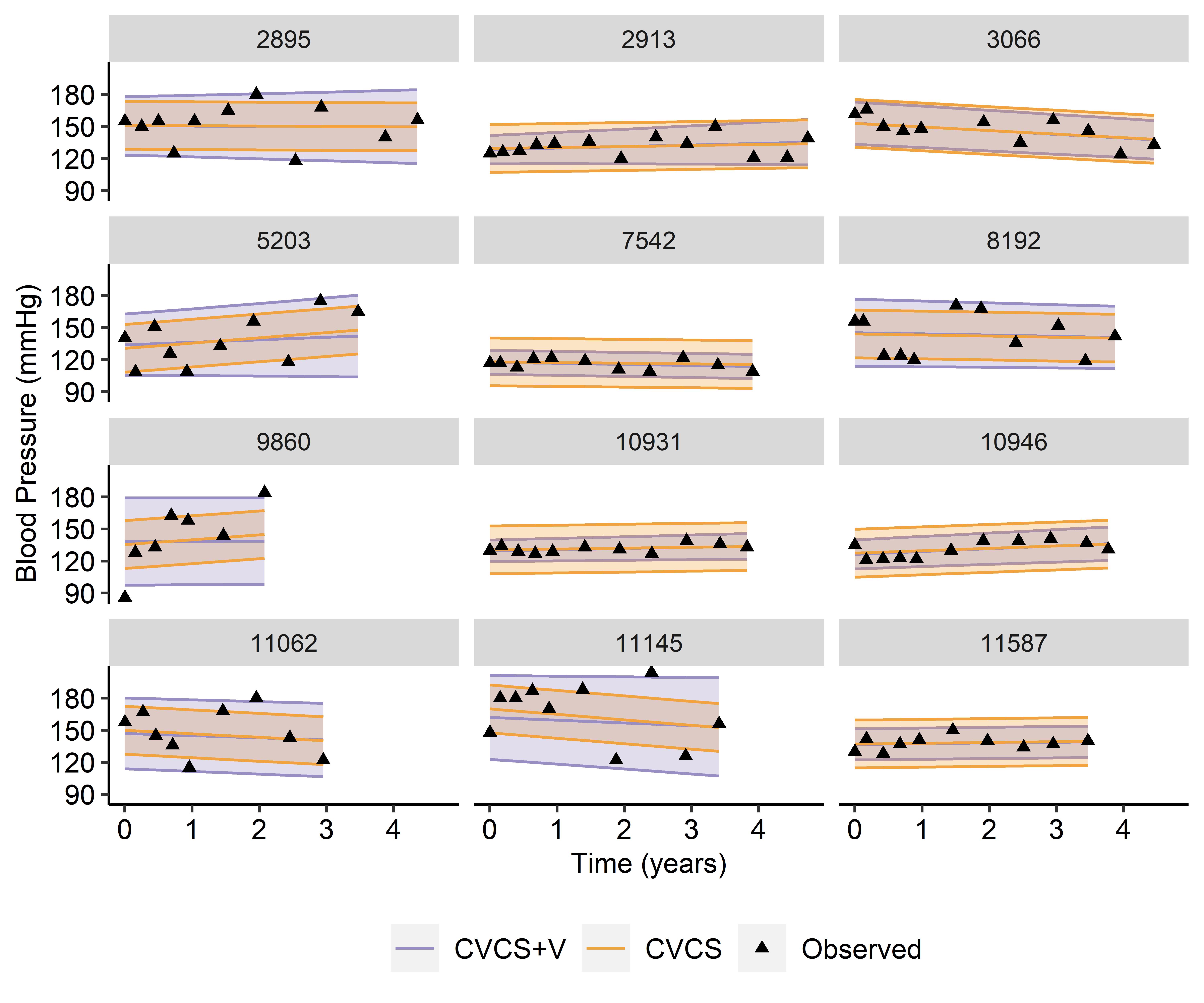}
    \caption{Prediction over time of the individual blood pressure and its prediction interval at 95\% for 6 subjects. Model CVCS+V assumed a time-dependent subject-specific variability and the model CVCS assumes a homogeneous and constant variability. The black triangles are the observed measurements.}
    \label{s:fit}
\end{figure}

\subsection{Predictions}

We compared the predictive abilities of models with and without time-dependent individual variability using AUC in a 5-fold cross-validation. The individual predictions of having CVD (or death) between 3 and 5 years for subjects free of any event at 3 years were computed using equation \eqref{pred_st}. The AUC was computed using the \texttt{timeROC} package \citep{blanche_2015}. The results are slightly better for the model with heterogeneous variability. We obtained respectively 0.609 (0.067) and 0.576 (0.067) for the risk of CVD, and 0.637 (0.078) and 0.616 (0.079) for the risk of death.\\

\begin{figure}[!ht]
    \centering
    \includegraphics{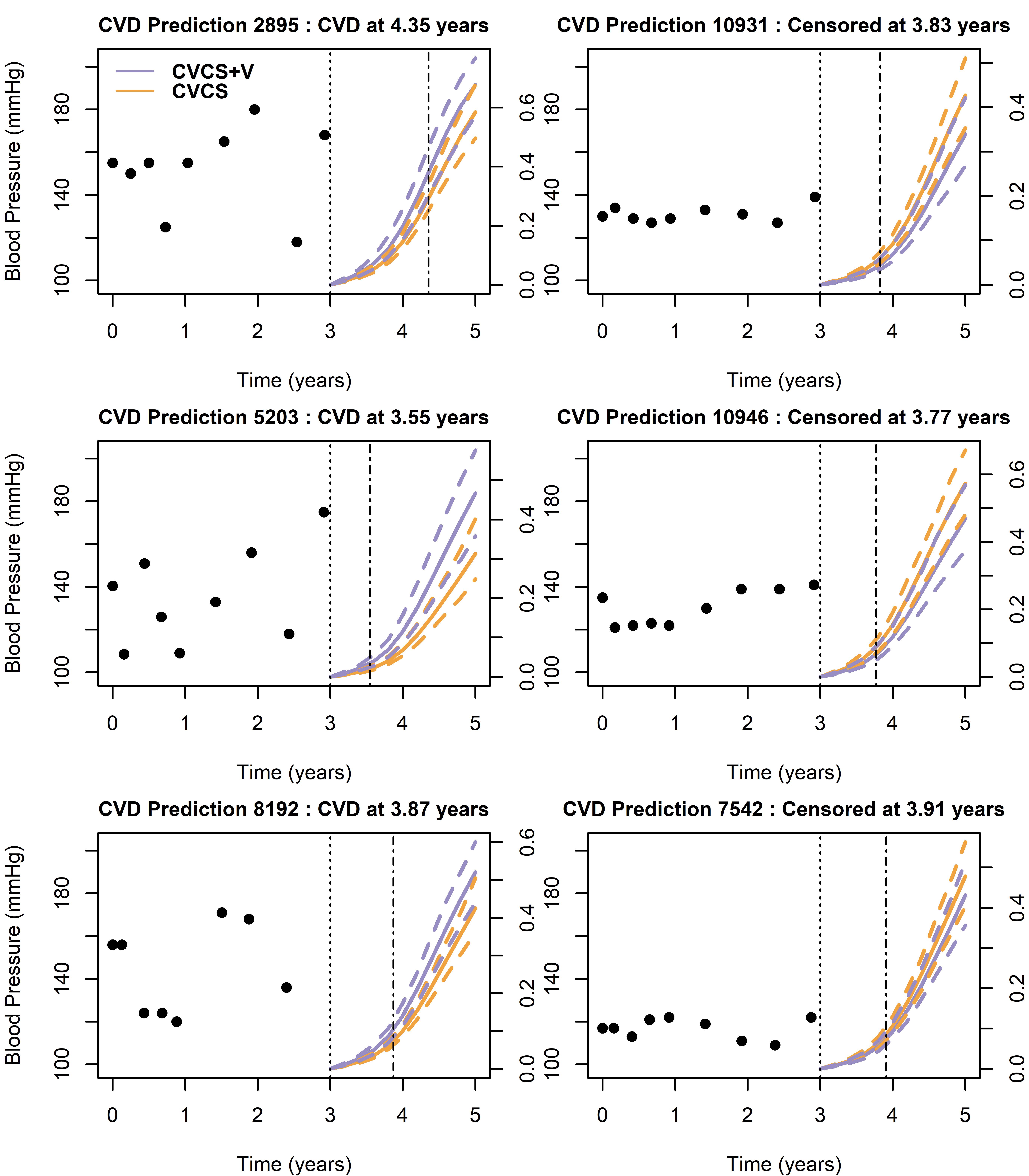}
    \caption{Prediction of the risk of CVD between 3 and 5 years (with its $95\%$ confidence interval indicated by dashed lines), for six patients at risk at 3 years, for Model CVCS+V (purple) and Model CVCS (orange). The dashed lines represent the observed time.}
    \label{s:predCVD}
\end{figure}

\begin{figure}[!ht]
    \centering
    \includegraphics{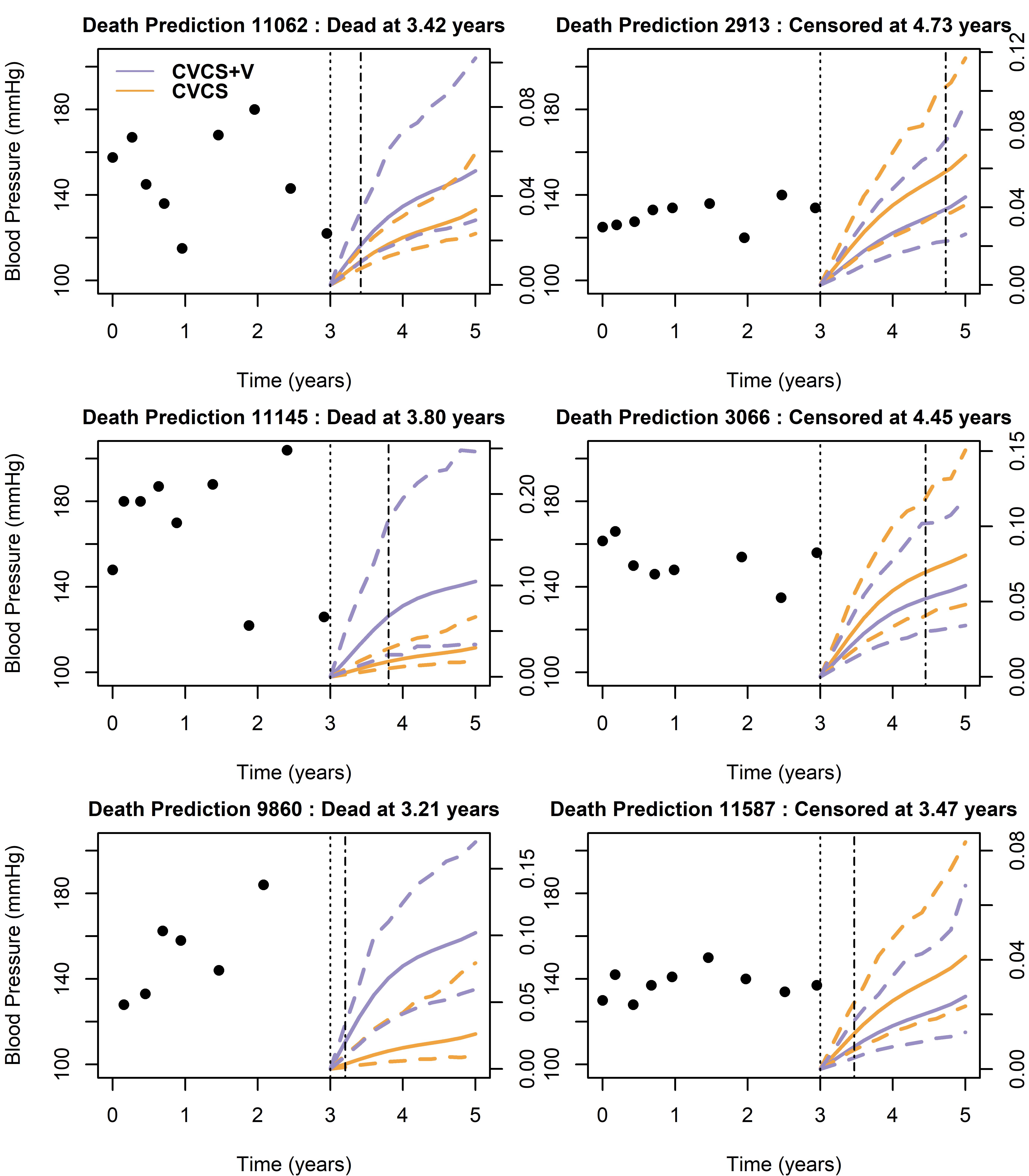}
    \caption{Prediction of the risk of Death between 3 and 5 years (with its $95\%$ confidence interval indicated by dashed lines), for six patients at risk at 3 years, for Model CVCS+V (purple) and Model CVCS (orange). The dashed lines represent the observed time.}
    \label{s:predDeath}
\end{figure}

To illustrate the effect of taking into account the current value of individual variance, we also computed the predicted risk of the events between 3 and 5 years for different subjects from both models, with and without time-dependent individual variability. We used the subjects selected for Figure \ref{s:fit} and present their predictions obtained via the cross-validation procedure\color{black}. Figures \ref{s:predCVD} and \ref{s:predDeath} shows that, for both the risk of CVD and the risk of death, the prediction is higher with the complete model when the individual experienced the event between 3 and 5 years than with the model without the heterogeneous variability. More, the predicted risk is smaller with the complete model when the individual do not experience the corresponding event.\\

\section{Discussion}
\label{s:discuss}

In this work, we have proposed a new joint model with a subject-specific time-dependent variance that extends the models proposed by Gao et al. \citep{gao_2011} and Barrett et al. \citep{barrett_2019}. Indeed, this new model allows time and covariate dependent individual variance and a flexible dependence structure between the competing events and the longitudinal marker. In particular, the risk of events may depend on both the current value and the current slope of the marker, in addition to the subject-specific time-dependent standard deviation of the residual error. This is an important asset of the model given that, in most health research contexts, it is more sensible to assume that the event risk depends on the time-dependent current value or slope of the marker instead of only time-independent random effects. Moreover, accounting for competing events may be important in many clinical applications. Simulation study allows us to demonstrate the good performance of the estimation procedure and to study the impact of the choice of $S$1 and $S2$. The model converged without bias and with good coverage rates, whatever the number of individual and the number of visits. Moreover, the estimates of the time-to-event sub-model are quite robust to a misspecification of the marker trajectory. In addition, we provided an R-package that allows frequentist estimation with a robust estimation algorithm which had shown very good behaviour in our simulations and in a previous work with different models \citep{philipps_2021}.\\
The analysis of the PROGRESS trial has shown that a high variability of blood pressure is associated with a high risk of CVD and death from other causes. Moreover, the individual residual variability depends on treatment group. These results are difficult to generalise to the entire population as the population study considered in this clinical trial is for the secondary prevention of stroke.\\
In this work, we have supposed that the visit times were not informative and that missing measurements before the event were missing at random. In the PROGRESS clinical study, these hypotheses are quite plausible since visits were planed following a pre-specified protocol and the rate of missed visits before the event was low (less than 3\%). For application to observational studies, it could be useful to extend this approach to consider an informative observation process. However, such a model would require three submodels: a mixed model for the evolution of the marker, a submodel for repeated events to describe the visit process and a model for the competing events of interest. This model would rely on non-verifiable parametric assumptions and its estimation process would be much more cumbersome.\\
The proposed approach addressed both right censoring and left-truncation, the two most common observation schemes for time-to-event data. Considering interval censoring and semi-competing events could represent a valuable enhancement. This extension would be useful when the exact time of onset of the main event is unknown (dementia for instance) and the competing event may arise after the main event (death). However, this would necessitate modeling the three transition intensities and the interval censoring would significantly complicate the computation of the likelihood.
\\
Such joint models with dependence on the heterogeneous variance (that can be viewed as an extension of the location-scale mixed model \citep{hedecker_2013}) are of great interest to investigate the association between the variability of markers or risk factors and the risk of health events in various fields of medical research, possibly allowing to improve the prediction ability for the event. For instance, hypotheses have emerged about the link between emotional instability and the risk of psychiatric events, or the variability of glycemia and the prognosis of diabetes. Thanks to wearable devices, recent medical research studies often include frequent repeated measures of exposures or biomarkers, allowing the investigation of hypotheses regarding the variability.

\section*{Acknowledgments}
Computer time for this article was provided by the computing facilities MCIA of the Université de Bordeaux and of the Université de Pau et des Pays de l'Adour.

\section*{Declaration of conflicting interests}
The authors declared no potential conflicts of interest with respect to the research, authorship, and/or publication of this article.

\section*{Funding}
This work was funded by the French National Research Agency (grant ANR-21-CE36 for the project "Joint Models for Epidemiology and Clinical research").\\
This PhD program is supported within the framework of the PIA3 (Investment for the Future). Project reference: 17-EURE-0019.\\

\bibliographystyle{unsrtnat}
\bibliography{arxiv.bib}
\newpage
\begin{center}
\LARGE{Supplementary Material \\}
\LARGE{A location-scale joint model for studying the link between the time-dependent subject-specific variability of blood pressure and competing events\\}
\vspace{3mm}
\normalsize
Léonie Courcoul$^{1*}$, Christophe Tzourio$^1$, Mark Woodward$^{2,3}$,\\
Antoine Barbieri$^1$,  and Hélène Jacqmin-Gadda$^1$

\noindent$^1$Univ. Bordeaux, INSERM, Bordeaux Population Health, U1219, France\\
$^2$ The George Institute for Global Health, Imperial College London, UK\\
$^3$ The George Institute for Global Health, University of New South Wales, Sydney, Australia\\
\vspace{2mm}
\end{center}
\beginsupplement
\begin{table}[h!]
\centering
\caption{Simulation results for scenario A with 1000 subjects (7 measures, $b_i$ and $\tau_i$ independent).*}
\label{t:SA1000}
\scalebox{0.7}{
\begin{tabular}{|lll|cccc|cccc|}
\hline
\multicolumn{3}{|c|}{\textbf{Parameter}}                                                                                                   & \multicolumn{4}{c|}{\textbf{Step 1}}                                                                    & \multicolumn{4}{c|}{\textbf{Step 2}}                                                                     \\ \hline
\multicolumn{2}{|l}{\textit{\textbf{}}}                                                              & \multicolumn{1}{l|}{\textbf{True}}  & \textbf{Mean} & \textbf{Empirical} & \textbf{Mean asymptotic} & \multicolumn{1}{l|}{\textbf{Coverage}}  & \textbf{Mean} & \textbf{Empirical} & \textbf{Mean asymptotic} & \multicolumn{1}{l|}{\textbf{Coverage}}  \\
\multicolumn{2}{|l}{\textit{\textbf{}}}                                                              & \multicolumn{1}{l|}{\textbf{value}} & \textbf{}     & \textbf{SE}        & \textbf{SE}              & \multicolumn{1}{l|}{\textbf{rate ($\%$)}} & \textbf{}     & \textbf{SE}        & \textbf{SE}              & \multicolumn{1}{l|}{\textbf{rate ($\%$)}} \\ \hline
\multicolumn{11}{|l|}{Longitudinal submodel}                                                                                                                                                                                                                                                                                                                   \\ \hline
\textit{Intercept}         & $\beta_0 $                                                & \multicolumn{1}{l|}{142}            & 142.0       & 0.533              & 0.509                    & \multicolumn{1}{l|}{94.97}             & 142.0       & 0.530              & 0.515                    & \multicolumn{1}{l|}{95.33}             \\
\hline
\end{tabular}}

\vspace{3mm}
SE: Standard Error; Coverage rate: coverage rate of the $95\%$ confidence interval.\\
* Results for 298 replicates with complete convergence over 300 for step 1 and 300 replicates over 300 for step 2.
\end{table}


\begin{table}[h!]
\caption{Simulation results for scenario B with 1000 subjects (13 measures, $b_i$ and $\tau_i$ independent).*}
\label{t:SB1000}
\scalebox{0.7}{\begin{tabular}{|lllcccccccc|}
\hline
\multicolumn{3}{|c|}{\textbf{Parameter}}                                                                                                   & \multicolumn{4}{c|}{\textbf{Step 1}}                                                                                                                                & \multicolumn{4}{c|}{\textbf{Step 2}}                                                                                                                                \\ \hline
\multicolumn{2}{|l}{\textit{\textbf{}}}                                                              & \multicolumn{1}{l|}{\textbf{True}}  & \multicolumn{1}{l}{\textbf{Mean}} & \multicolumn{1}{l}{\textbf{Empirical}} & \multicolumn{1}{l}{\textbf{Mean asymptotic}} & \multicolumn{1}{l|}{\textbf{Coverage}}  & \multicolumn{1}{l}{\textbf{Mean}} & \multicolumn{1}{l}{\textbf{Empirical}} & \multicolumn{1}{l}{\textbf{Mean asymptotic}} & \multicolumn{1}{l|}{\textbf{Coverage}}  \\
\multicolumn{2}{|l}{\textit{\textbf{}}}                                                              & \multicolumn{1}{l|}{\textbf{value}} & \multicolumn{1}{l}{\textbf{}}     & \multicolumn{1}{l}{\textbf{SE}}        & \multicolumn{1}{l}{\textbf{SE}}              & \multicolumn{1}{l|}{\textbf{rate ($\%$)}} & \multicolumn{1}{l}{\textbf{}}     & \multicolumn{1}{l}{\textbf{SE}}        & \multicolumn{1}{l}{\textbf{SE}}              & \multicolumn{1}{l|}{\textbf{rate ($\%$)}} \\ \hline
\multicolumn{11}{|l|}{Longitudinal submodel}                                                                                                                                                                                                                                                                                                                                                                                                                                           \\
\hline
\end{tabular}}

\vspace{3mm}
\noindent SE: Standard Error; Coverage rate: coverage rate of the $95\%$ confidence interval.\\
* Results for 299 replicates with complete convergence over 300 for step 1 and 300 replicates over 300 for step 2.
\end{table}

\begin{table}[]
\caption{Simulation results for scenario C with 1000 subjects (7 measures, $b_i$ and $\tau_i$ correlated).*}
\label{t:SC1000}
\scalebox{0.7}{
\begin{tabular}{|lllcccccccc|}
\hline
\multicolumn{3}{|c|}{\textbf{Parameter}}                                                                                                                        & \multicolumn{4}{c|}{\textbf{Step 1}}                                                                                                                                & \multicolumn{4}{c|}{\textbf{Step 2}}                                                                                                                                \\ \hline
\multicolumn{2}{|l}{\textit{\textbf{}}}                                                                                   & \multicolumn{1}{l|}{\textbf{True}}  & \multicolumn{1}{l}{\textbf{Mean}} & \multicolumn{1}{l}{\textbf{Empirical}} & \multicolumn{1}{l}{\textbf{Mean asymptotic}} & \multicolumn{1}{l|}{\textbf{Coverage}}  & \multicolumn{1}{l}{\textbf{Mean}} & \multicolumn{1}{l}{\textbf{Empirical}} & \multicolumn{1}{l}{\textbf{Mean asymptotic}} & \multicolumn{1}{l|}{\textbf{Coverage}}  \\
\multicolumn{2}{|l}{\textit{\textbf{}}}                                                                                   & \multicolumn{1}{l|}{\textbf{value}} & \multicolumn{1}{l}{\textbf{}}     & \multicolumn{1}{l}{\textbf{SE}}        & \multicolumn{1}{l}{\textbf{SE}}              & \multicolumn{1}{l|}{\textbf{rate ($\%$)}} & \multicolumn{1}{l}{\textbf{}}     & \multicolumn{1}{l}{\textbf{SE}}        & \multicolumn{1}{l}{\textbf{SE}}              & \multicolumn{1}{l|}{\textbf{rate ($\%$)}} \\
\hline 
\end{tabular}}

\vspace{3mm}
SE: Standard Error; Coverage rate: coverage rate of the $95\%$ confidence interval.\\
* Results for 299 replicates with complete convergence over 300 for step 1 and 300 replicates over 300 for step 2.

\end{table}

\begin{table}[]
\caption{Simulation results for scenario D with 500 subjects (13 measures, $b_i$ and $\tau_i$ correlated).*}
\label{t:SD500}
\scalebox{0.7}{
\begin{tabular}{|lllcccccccc|}
\hline
\multicolumn{3}{|c|}{\textbf{Parameter}}                                                & \multicolumn{4}{c|}{\textbf{Step 1}}                                                                                                                               & \multicolumn{4}{c|}{\textbf{Step 2}}                                                                                                                               \\ \hline
\multicolumn{2}{|l}{\textit{\textbf{}}}           & \multicolumn{1}{l|}{\textbf{True}}  & \multicolumn{1}{l}{\textbf{Mean}} & \multicolumn{1}{l}{\textbf{Empirical}} & \multicolumn{1}{l}{\textbf{Mean asymptotic}} & \multicolumn{1}{l|}{\textbf{Coverage}} & \multicolumn{1}{l}{\textbf{Mean}} & \multicolumn{1}{l}{\textbf{Empirical}} & \multicolumn{1}{l}{\textbf{Mean asymptotic}} & \multicolumn{1}{l|}{\textbf{Coverage}} \\
\multicolumn{2}{|l}{\textit{\textbf{}}}           & \multicolumn{1}{l|}{\textbf{value}} & \multicolumn{1}{l}{\textbf{}}     & \multicolumn{1}{l}{\textbf{SE}}        & \multicolumn{1}{l}{\textbf{SE}}              & \multicolumn{1}{l|}{\textbf{rate (\%)}} & \multicolumn{1}{l}{\textbf{}}     & \multicolumn{1}{l}{\textbf{SE}}        & \multicolumn{1}{l}{\textbf{SE}}              & \multicolumn{1}{l|}{\textbf{rate (\%)}} \\ \hline
\multicolumn{11}{|l|}{Longitudinal submodel}                                                                                                                                                                                                                                                                                                                                                                                      \\
\hline
\end{tabular}}

\vspace{3mm}
SE: Standard Error; Coverage rate: coverage rate of the $95\%$ confidence interval.\\
* Results for 300 replicates with complete convergence over 300 for step 1 and 299 replicates over 300 for step 2.
\end{table}

\begin{table}[]
\caption{Simulation results for scenario D with 1000 subjects (13 measures, $b_i$ and $\tau_i$ correlated).*}
\label{t:SD1000}
\scalebox{0.7}{
\begin{tabular}{|lllcccccccc|}
\hline
\multicolumn{3}{|c|}{\textbf{Parameter}}                                                  & \multicolumn{4}{c|}{\textbf{Step 1}}                                                                                                                               & \multicolumn{4}{c|}{\textbf{Step 2}}                                                                                                                               \\ \hline
\multicolumn{2}{|l}{\textit{\textbf{}}}             & \multicolumn{1}{l|}{\textbf{True}}  & \multicolumn{1}{l}{\textbf{Mean}} & \multicolumn{1}{l}{\textbf{Empirical}} & \multicolumn{1}{l}{\textbf{Mean asymptotic}} & \multicolumn{1}{l|}{\textbf{Coverage}} & \multicolumn{1}{l}{\textbf{Mean}} & \multicolumn{1}{l}{\textbf{Empirical}} & \multicolumn{1}{l}{\textbf{Mean asymptotic}} & \multicolumn{1}{l|}{\textbf{Coverage}} \\
\multicolumn{2}{|l}{\textit{\textbf{}}}             & \multicolumn{1}{l|}{\textbf{value}} & \multicolumn{1}{l}{\textbf{}}     & \multicolumn{1}{l}{\textbf{SE}}        & \multicolumn{1}{l}{\textbf{SE}}              & \multicolumn{1}{l|}{\textbf{rate (\%)}} & \multicolumn{1}{l}{\textbf{}}     & \multicolumn{1}{l}{\textbf{SE}}        & \multicolumn{1}{l}{\textbf{SE}}              & \multicolumn{1}{l|}{\textbf{rate (\%)}} \\ \hline
\multicolumn{11}{|l|}{Longitudinal submodel}                                                                                                                                                                                                                                                                                                                                                                                        \\
\hline
\end{tabular}}

\vspace{3mm}
SE: Standard Error; Coverage rate: coverage rate of the $95\%$ confidence interval.\\
* Results for 300 replicates with complete convergence over 300.
\end{table}

\begin{table}[h!]
\caption{Simulation results for scenario E with 500 subjects (Misspecified model).*}
\label{t:SE500}
\scalebox{0.7}{\begin{tabular}{|lllcccccccc}
\hline
\multicolumn{3}{|c|}{\textbf{Parameter}}                                                                                           & \multicolumn{4}{c|}{\textbf{Step 1}}                                                                    & \multicolumn{4}{c}{\textbf{Step 2}}                                                                     \\ \hline
\multicolumn{2}{|l}{\textit{\textbf{}}}                                                              & \multicolumn{1}{l|}{True}   & \textbf{Mean} & \textbf{Empirical} & \textbf{Mean asymptotic} & \multicolumn{1}{c|}{\textbf{Coverage}}  & \textbf{Mean} & \textbf{Empirical} & \textbf{Mean asymptotic} & \multicolumn{1}{c|}{\textbf{Coverage}}  \\
\multicolumn{2}{|l}{\textit{\textbf{}}}                                                              & \multicolumn{1}{l|}{Value}  & \textbf{}     & \textbf{SE}        & \textbf{SE}              & \multicolumn{1}{c|}{\textbf{rate ($\%$)}} & \textbf{}     & \textbf{SE}        & \textbf{SE}              & \multicolumn{1}{c|}{\textbf{rate ($\%$)}} \\ \hline
\multicolumn{11}{|l|}{Longitudinal submodel}                                                                                                                                                                                                                                                                                                           \\ \hline
\textit{Intercept}         & $\beta_0$                                                 & \multicolumn{1}{l|}{142}    & 140.8        & 0.808               & 0.667                     & \multicolumn{1}{c|}{52.67}              & 140.8        & 0.791               & 0.701                     & \multicolumn{1}{c|}{56.67}              \\
\hline
\end{tabular}}
\vspace{3mm}\\
SE: Standard Error; Coverage rate: coverage rate of the $95\%$ confidence interval.\\
* Results for 300 replicates with complete convergence over 300 for step 1 and step 2.
\end{table}

\begin{table}[h!]
\centering
\caption{Parameter estimates of the CVCS joint model on the Progress clinical trial data.}
\label{t:CVCSres}
\begin{tabular}{|lccc|}
\hline
\textbf{Parameter} & \textbf{Estimate} & \textbf{Standard error} & \textbf{p-value} \\
  \hline

\multicolumn{4}{|l|}{\textit{Survival submodel for CVD}} \\ 
\hspace{6mm} BP current value & 0.003 & 0.003 & $0.318$\\
\hspace{6mm} BP current slope &  -0.122 & 0.017 & $<0.001$\\
\hspace{6mm} treatment group &  -0.093 & 0.081 & $0.253$\\
\hspace{6mm} male & 0.277 & 0.087 & $0.002$\\
\hspace{6mm} age & 0.041 & 0.005 & $<0.001$\\
\multicolumn{4}{|l|}{\textit{Survival submodel for Death}} \\
\hspace{6mm} BP current value & -0.005 & 0.007 & $0.487$\\
\hspace{6mm} BP current slope &  -0.100 & 0.045 & $0.038$\\
\hspace{6mm} treatment group &  -0.008 & 0.067 & $0.907$\\
\hspace{6mm} male & 0.493 & 0.194 & $0.011$\\
\hspace{6mm} age & 0.057 & 0.010 & $<0.001$\\
\multicolumn{4}{|l|}{\textit{Longitudinal submodel}} \\
\multicolumn{4}{|l|}{\underline{Blood Pressure Mean}} \\
\hspace{6mm} intercept & 142.7 & 0.345 & $<0.001$\\
\hspace{6mm} time & -0.146 & 0.073 & 0.046\\
\hspace{6mm} treatment group & -7.942 & 0.457 & $<0.001$\\
\hline
\end{tabular}\\
\vspace{5mm}
BP: Blood Pressure
\end{table}

\begin{figure}[!ht]
    \centering
    \includegraphics{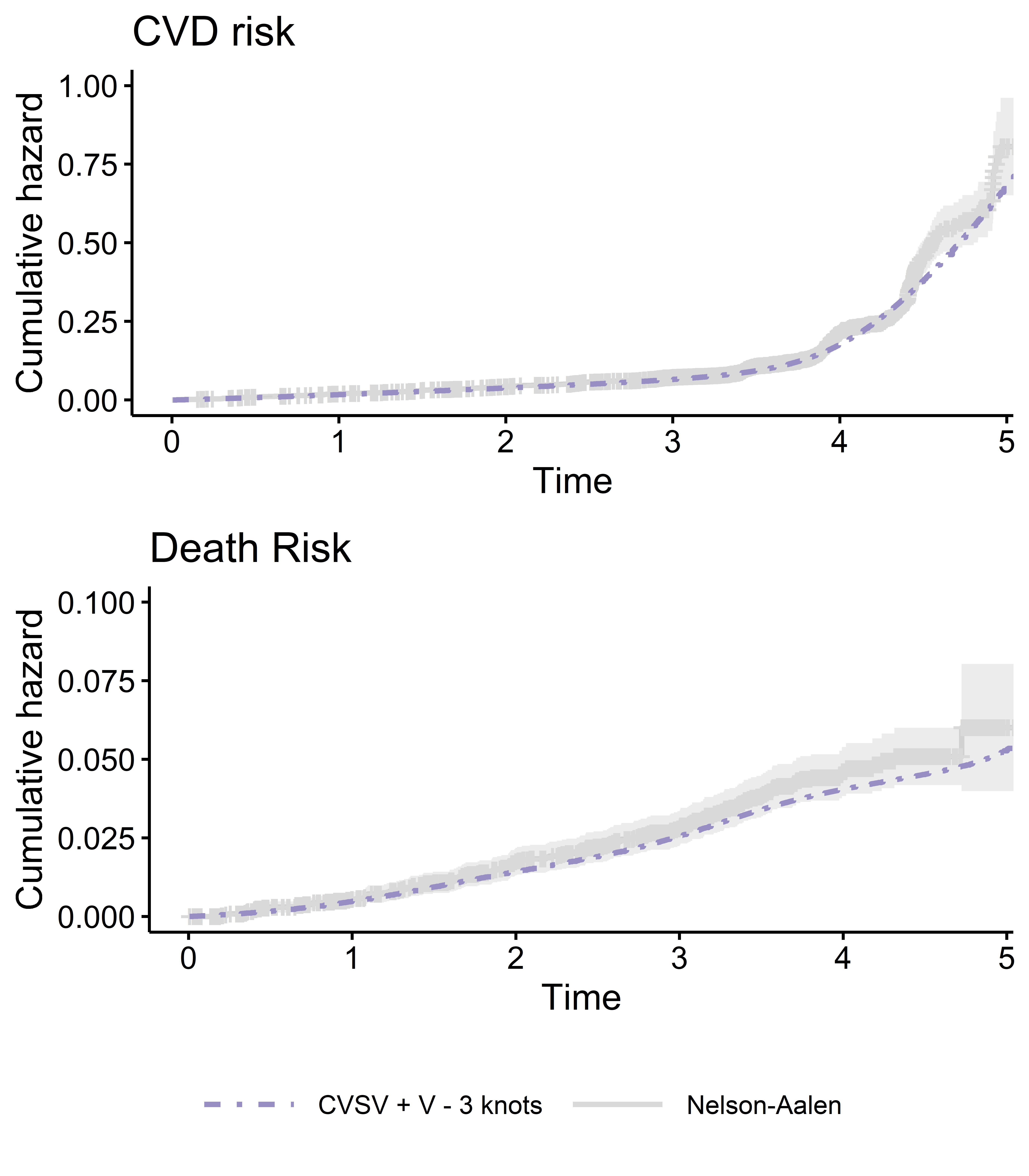}
    \caption{Survival submodel for CVD (top) and death (bottom) fit assessment: comparison between predicted cumulative hazard function (in purple) and Nelson Aalen estimator (in grey).}
    \label{F:FigSup1}
\end{figure}

\begin{figure}[!ht]
    \centering
    \includegraphics[scale=0.65]{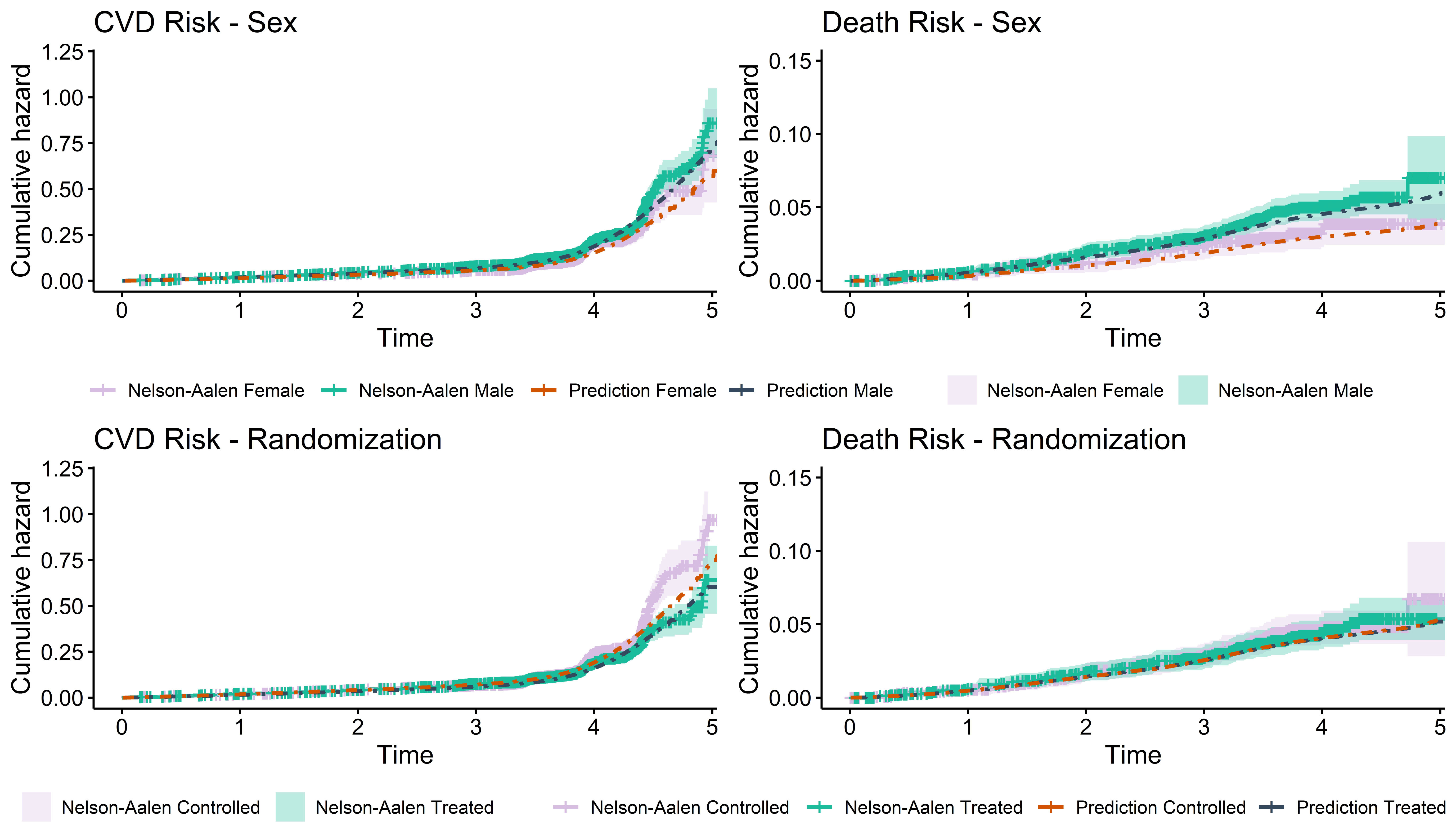}
    \caption{Survival submodel for CVD (left) and death (right) fit assessment for Sex (top) and Randomization group (bottom): comparison between predicted cumulative hazard function and Nelson Aalen estimator.}
    \label{F:FigSup2}
\end{figure}

\end{document}